\documentclass[12pt]{article}
\usepackage{amsmath, amssymb}
\usepackage{graphicx,psfrag,epsf}
\usepackage{enumerate}
\usepackage{natbib}
\usepackage{bbm}
\usepackage{url} 

\newcommand{\blind}{1}

\newcommand{\Lnet}{\mathfrak{L}}
\newcommand{\T}{^{\mathsf T}}

\DeclareMathOperator{\nach}{ne}

\DeclareMathOperator{\pa}{pa}

\DeclareMathOperator{\child}{ch}

\DeclareMathOperator{\dg}{deg_{\mathcal{G}}}

\begin{document}

\def\spacingset#1{\renewcommand{\baselinestretch}%
{#1}\small\normalsize} \spacingset{1}


\if1\blind
{
  \title{\bf Structured network regression for spatial point patterns}
  \author{Matthias Eckardt\thanks{
    The authors gratefully acknowledge financial support from the Spanish Ministry of Economy and Competitiveness via grant MTM 2013-
43917-P}\hspace{.2cm}\\
    Department of Computer Science, Humboldt Universit\"{a}t zu Berlin, Germany\\
    and \\
   Jorge Mateu\\
    Department of Mathematics, University Jaume I, Castell\'{o}n, Spain}
  \maketitle
} \fi

\if0\blind
{
  \bigskip
  \bigskip
  \bigskip
  \begin{center}
    {\LARGE\bf Structured network regression for spatial point patterns}
\end{center}
  \medskip
} \fi

\bigskip
\begin{abstract}
The analysis of spatial point patterns that occur in the network domain have recently gained much attraction and various intensity functions and measures have been proposed. However, the linkage of spatial network statistics to regression models has not been approached so far. 
This paper presents a new regression approach which treats a generic intensity function of a planar point pattern that occurred on a network as the outcome of a set of different covariates and various graph statistics. Different to all alternative approaches, our model is the first which permits the statistical analysis of complex regression data in the context of network intensity functions for spatial point patterns. The potential of our new technique to model the structural dependencies of network intensity functions on various covariates and graph statistics is illustrated using call-in data on neighbour and community disturbances in an urban context.   
\end{abstract}

\noindent%
{\it Keywords:} Structured additive effects, Spatial network  model, Spatial point pattern
\vfill

\newpage
\spacingset{1.45} 

\section{Introduction}\label{sec:int}

The analysis of complex structures has become a highly attractive field in methodological as well as applied research within different disciplines. Here, one very prominent topic are relationships that are present in network structures.  Most generally, the objective of interest here lies on the investigation of structures between different entities of the network. For an intense description of different aspects of network analysis we refer the interested reader to \cite{Carrington2005}, \cite{Kolaczyk:2009} and \cite{Goldenberg:EtAl:2010}.

Although most of this research has been conducted within the field of social network analysis or statistical physics, a growing number of papers considers network structures in the context of spatial processes including the investigation of spatial point patterns. The statistical analysis of such spatial point pattern data that occurs on linear networks has been pioneered by \cite{Okabe:Yamada:2001}. This paper has later been extended by \cite{Ang:2010}, \cite{Ang:Baddeley:Nair2012} as well as \cite{Baddeley:Jammalamadaka:Nair:2014} who derived a geometrically-adjusted extension of Ripleys' $K$-function \citep{ripley:76} to planar point processes on linear networks. As the underlying assumption, these models treat a point pattern $x$ on a linear network $\Lnet$ as a realization of a point process that occurs randomly on a linear network in a bounded region on the linear network space $\Lnet\subset\mathbb{R}$. Further contributions with respect to this linear network formalism are presented by \cite{Okabe:Satoh:2009},  \cite{Okabe:Sugihara:2012} and \cite{Borruso:2005,Borruso:2008}. 
 
An alternative approach for the analysis of spatial point patterns that is not restricted to the linear network formalism or to simple point processes has recently been introduced by \cite{Eckardt:Mateu2016}. This alternative formalism allows to take undirected, directed or partially directed graph structures  as well as temporal dynamics into consideration. In contrast to \cite{Ang:2010} or \cite{Ang:Baddeley:Nair2012}, this alternative approach highlights the possibility to achieve several different graph-based intensity formulations and related statistics for network structures. Different from the linear network formalism, \citet{Eckardt:Mateu2016} assume that a point pattern appears randomly between pairs of geo-referenced nodes whose location within a planar region is treated as fixed. In this respect, the intensity of a process appears as an edgewise counting measure adjusted for the geodesic distance between the endpoints of the edge. 

This leads to the non-circular definition of three different classes of  intensity measures, and finally to different versions of Ripleys' $K$-function based on subsets of adjacent network segments. The first class contains various intensity measures and related statistics that are calculated with respect to distinct edges. In contrast to this edgewise calculation, all measures can be formulated with respect to sequences of distinct nodes and distinct edges leading to pathwise statements. This pathwise perspective forms the second class of intensity functions.  Lastly, the third class of such measures consists of average nodewise intensities which are obtained as an average intensity over subsets of adjacent nodes. These nodewise intensities are computed as means over the different sets of edge intensity functions. 

Although the results of \citet{Eckardt:Mateu2016} allow for deeper insights in the behavior of possibly marked point patterns and the relational information contained in the network structure, the possible impact of additional covariates is not yet captured. Principally, this limitation might affect all three classes of intensity measures: the edgewise, the pathwise and the nodewise mean intensity function. Here, possible influences by different types of covariates seem to be a plausible assumption. To approach this limitation, we treat a generic network intensity function $\lambda(\phi) $ as the conditional expectation of a point pattern $\phi$ over a set of fixed structural elements of a network within a structured network regression model. These network elements include distinct edges for edgewise intensities, paths for pathwise intensities, and nodes in case of nodewise mean intensity functions.      

This paper is in general the first which relates the intensity function to a regression model within the context of network structures where we assume that the joint distribution of a point pattern $\phi$ associated with a specific set of network elements given a set of explanatory variables and parameters belongs to an exponential family. This formulation is different from exponential random graph models (ERGMs) in which the adjacency of a network is treated as the outcome of a regression. In this sense, ERGMs aim to explore the effects on the network structure. For a detailed discussion of ERGMs we refer the interested reader to \cite{Koskinen:Lusher:Robins:2011}. Differently from ERGMs, our focus here lies on the effects on structurally calculated generic intensity functions where the network structure is treated as fixed.

This paper is organized as follows: the main results of \citet{Eckardt:Mateu2016} and  different possible network intensity functions are described in Section \ref{sec:prior}. Hereafter, Section \ref{sec:reg} introduces a generalized structured regression model for network mean intensity functions. Applications of the regression model to neighbor and community disturbances data are discussed in Section \ref{sec:apl}. Finally, the concluding Section \ref{sec:conclusion} comments on the major results and impacts on future research. 

\section{Network intensity functions for spatial point patterns }\label{sec:prior}

This section focuses on the introduction of different edgewise and averaged nodewise intensity functions as proposed in \cite{Eckardt:Mateu2016} and only briefly covers the underlying graph theoretical concepts. Pathwise intensity functions which are generalizations of edgewise intensities to sequences of distinct nodes and edges will not explicitly be covered here. For a comprehensive treatment of graph theory and further network related intensity functions as well as a linkage of graphical modeling to planar point processes we refer to \cite{Eckardt:2016} and \citet{Eckardt:Mateu2016} and the references therein.

Formally, a network is expressed in terms of a graph $G=(V,E)$ in which every node $v_i(s_{v_i})$ is indexed with a pair of fixed coordinates $s_{v_i}=(x_{v_i},y_{v_i})$ encoding locations of interest such as crossings in a traffic  network. These fixed positions are pairwise joint by edges which are treated as edge intervals $s_{e_i}=(s_{v_i},s_{v_j})$ of arbitrary length that are spanned between two nodes. In this context, \citet{Eckardt:Mateu2016} define the set of all $k$ edge intervals in the graph as $\mathcal{S}_{E(G)}=\lbrace s_{e_1},\ldots,s_{e_k}\rbrace$. The realization of a point process $X$ with respect to a network is then understood as a random event that occurs on a location within a closed interval belonging to $\mathcal{S}_{E(G)}$. This random location is formally expressed as $\tilde{s}=(\tilde{x}, \tilde{y})$. 

The number of points that fall into the edges could then be expressed by a counting measure $N(\cdot)$. If the graph only consists of undirected edges, this counting measure is defined as 
\[
N(s_{e_i})=\sum\mathbbm{1}_{\lbrace x_{v_i}\leq \tilde{x}\leq x_{v_j},y_{v_i}\leq \tilde{y}\leq y_{v_j}\rbrace}X(\tilde{s}), x_{v_i}<x_{v_j},y_{v_i}<y_{v_j}.
\]

To define nodewise mean intensity functions, we first introduce a pairwise intensity function 
\[
\lambda(s_{e_i})=\lim_{|ds_{e_i}|\rightarrow 0}\left\{\frac{\mathbbm{E}\left[N(ds_{e_i})\right]}{|ds_{e_i}|}\right\}, s_{e_i}\in\mathcal{S}_{E(G)}
\] 
where $ds_{e_i}$ is an infinitesimal interval. From this, an intensity related to neighboring vertices is obtained as  
\[
\lambda(v_i)=\frac{1}{|\dg(v_i)|}\sum_{v_j\in\nach(v_i)}\lambda(s_{e_i}). 
\]
where $ e_i=(v_i,v_j),\nach(v_i)$ is the set of all $k$-nearest neighboring nodes connected to $v_i$ and $\dg(v_i)$ is the size of $\nach(v_i)$ .

Similar intensity functions for alternative graph structures can be achieved as modification of the previous definitions. For a graph which only consists of directed edges two possible counting measures exist. These are, 
\[
N(s_{e_i}^{in})=\sum\mathbbm{1}_{\lbrace\pa(x_{v_i}\leq \tilde{x}\leq x_{v_j},y_{v_i}\leq \tilde{y}\leq y_{v_j})\rbrace}X(\tilde{s}), x_{v_i}<x_{v_j}, y_{v_i}<y_{v_j}
\]
and 
\[
N(s_{e_i}^{out})=\sum\mathbbm{1}_{\lbrace\child(x_{v_i}\leq \tilde{x}\leq x_{v_j},y_{v_i}\leq \tilde{y}\leq y_{v_j})\rbrace}X(\tilde{s}), x_{v_i}<x_{v_j}, y_{v_i}<y_{v_j}
\]
where $\pa(x_{v_i})$ are all the edges pointing to $v_i$ and $\child(x_{v_i})$ are all the edges departing from $v_i$. Again, these counting measures can be used to define two types of directed nodewise mean intensities functions. For the set of edges which point to  $v_i$ this leads to 
\[
\lambda^{in}(v_i)=\frac{1}{|\dg^+(v_i)|}\sum_{v_j\in\pa(v_i)}\lambda(s_{e_i}^{in}) 
\]
where 
\[
\lambda(s_{e_i}^{in})=\lim_{|ds_{e_i}^{in}|\rightarrow 0}\left\{\frac{\mathbbm{E}\left[N(ds_{e_i}^{in})\right]}{|ds_{e_i}^{in}|}\right\}, s_{e_i}^{in}\in\mathcal{S}_{E(G)}
\] 
and $\dg^+(v_i)$ is the size of $\pa(x_{v_i})$. The mean related to the opposite relation, the set of arrows which point from $v_i$, can be expressed as 
\[
\lambda(s_{e_i}^{out})=\lim_{|ds_{e_i}^{out}|\rightarrow 0}\left\{\frac{\mathbbm{E}\left[N(ds_{e_i}^{out})\right]}{|ds_{e_i}^{out}|}\right\}, s_{e_i}^{out}\in\mathcal{S}_{E(G)}.
\] 
where 
\[
\lambda^{out}(v_i)=\frac{1}{|\dg^-(v_i)|}\sum_{v_j\in\child(v_i)}\lambda(s_{e_i}^{out}).
\]
and $\dg^-(v_i)$ is the  size of $\child(x_{v_i})$.

Besides intensity measures related to networks which only consist of directed or undirected edges, a third graph configuration might be of interest where directed and undirected edges appear simultaneously. In this case, all previously described counting measures and related statistics generally remain applicable with respect to subsets of distinct edges. Besides, additional measures could be derived as a combination of undirected and directed versions. In this respect, \citet{Eckardt:Mateu2016} presented an extended intensity measure related to the union of undirected and directed edges as
\[
\lambda^{cg}(v_i)=\frac{1}{|\dg^{cg}(v_i)|} \lambda^{out}(v_i)\cup \lambda^{in}(v_i)\cup \lambda(v_i).
\]
Alternatively, several re-definitions of $\lambda^{cg}(v_i)$ can be considered which only take certain unions of distinct subsets of edges such as $\pa(\cdot)\cup\child(\cdot)$  or  $\nach(\cdot)\cup\child(\cdot)$ into account.

\section{Structured network regression model}\label{sec:reg}

Besides the edgewise, pathwise or mean nodewise intensity functions of point processes that occur on differently shaped network graphs, we now consider the situation where we want to estimate or predict the edgewise, pathwise or mean nodewise intensity functions based on additional information. This additional information is treated as a generic list of $l$ exploratory variables including various network statistics as well as additional covariates of different type and different dimension. 

To begin, we assume that $\phi_i$ is the realization of a planar point process $\Phi_i$ related to the $i$-th structural element, $i=1,\ldots,n$, of a network. These structural elements could be edges, paths or nodes of a undirected, directed or partly directed network. Additionally, we denote a generic network intensity function related to the distinct types of $\lambda_i$ as introduced in \citet{Eckardt:Mateu2016} as $\lambda_i(\phi)$. That is, $\lambda_i(\phi)$ could either be related to edges, paths or nodes in undirected, directed  or partially directed networks as described in Section \ref{sec:prior}. 

The general idea which will be elaborated in this section is to write $\lambda_i(\phi)$ as the outcome of additive combinations of structural covariates. A principle regression model in this spirit is the generalized linear model as introduced by \cite{McCullagh1989}. Here, the distribution of an observation $\phi_i$, given a set of covariates $\mathbf{z}_i$ and unknown parameters $\gamma_i$, is assumed to belong to an exponential family 
\begin{equation}
f_{\Phi_i}(\phi_i|\vartheta_i,\psi)=\exp\left(\frac{\phi_i\vartheta_i-b(\vartheta_i)}{a_i(\psi)}+c(\phi_i,~\psi)\right),~i=1,\dots,n 
\label{eq:expofam}
\end{equation}
where $\psi$ is a scale parameter and $a(\cdot), b(\cdot)$ and $c(\cdot)$ are unknown functions. The conditional expectation $\mathbb{E}\left[\phi_i|\mathbf{z}_i, \gamma_i\right]=\mu_i$ is modeled by $g(\mu_i)=\eta_i$ or $\mu_i=h(\eta_i)$ where $g$ (resp. $h$) is a known link function (resp. response function) and $\eta_i=\mathbf{z}_i\T\gamma_i$ is a linear predictor. 

For our purpose, this linear predictor $\eta_i$ seems to be too restrictive and unable to capture all, possibly nonlinear, effects including for example graph statistics as well as spatial or temporal correlation among observations, and heterogeneity. A more flexible regression framework which unifies several  extensions of the generalized linear model such as the generalized additive model of \cite{Hastie1990} or the geoadditive model of \cite{Kammann2003} is the structured additive regression (STAR) model. In this model, the linear predictor $\eta_i$ is replaced by a structured additive predictor $\eta^{\star}_i=\sum^p_{j=1} f_j(\nu_{ij})+\mathbf{z}_i\T\gamma_i$ where $f_j$ are not necessary smooth functions of generic covariates $\nu_j$ of different type and dimension. To ensure the identifiability of $f_j(\cdot)$, the functions  are constraint to have a zero mean. For a detailed discussion of the structured additive formalism in general we refer the interested reader to \cite{Fahrmeir2004} and \cite{Brezger2006} and the references therein. Structured additive regression models for count data have been presented in \cite{Fahrmeir2003, Fahrmeir2006} and, focussing on geoadditive survival models, in \cite{Hennerfeld2006}.

To model a generic network intensity function $\lambda_i(\phi)$ related to the $i$-th network element, we integrate a set of fixed graph statistics $\mathbf{w}_i$ into the structured additive predictor. In more detail, setting $\mu_i=\lambda_i(\phi)$ we assume that the generic network intensity function is linked to a structured network predictor $\eta^{\circ}_i=\beta_0+\eta^{\star}_i+\mathbf{w}_i\T\xi_i$ where $\mathbf{w}_i$ is a set of graph statistics, $\xi_i$ is a set of unknown parameters and  $\eta^{\star}_i$ is a generic structured additive predictor related to the $i$-th network element. 
Precisely, the generic structured network predictor for the $i$-th structural network element is defined as 
\begin{equation}
\eta^{\circ}_i=\beta_0+\sum^p_{j=1} f_j(\nu_{ij})+f(\alpha_s)+\mathbf{z}_i\T\gamma_i+\mathbf{w}_i\T\xi_i
\label{Eq:SNR}
\end{equation}  
where $\beta_0$ represents a possible offset parameter, $f_j(\nu_{ij})$ and $\mathbf{z}_i\T\gamma_i$ model the  nonlinear and fixed effects aggregated at the $i$-th structural network level and $f(\alpha_s)$ encode coarser information recorded at a spatial lattice data level. Thus, apart from the spatial lattice information, all information collected in the structured network predictor is recorded at an identical level of structural network elements as $\lambda_i(\phi)$. The class of model with a predictor in form of \eqref{Eq:SNR} is called {\em{structured network regression model}}. Using a log-link yields to 
\[
\lambda_i(\phi)=\exp(\eta^{\circ}_i).
\]

Inference for the structured network regression is carried out using an empirical Bayesian framework. Here, the unknown functions $f_j(\nu_{ij})$ and the spatial lattice information $f(\alpha_i)$ as well as the fixed parameters $\gamma$ and graph statistics $\xi$ in \eqref{Eq:SNR} are treated as random and are supplemented by priors.
 
For the fixed linear effects $\gamma$ and the graph statistics $\xi$ we consider flat priors such that $p(\gamma)$ and $p(\xi)$ are proportional to a constant $c$. For the unknown functions, the prior choice depends on the type of covariate $\nu_i$ and on smoothness assumptions. To define a general form of the prior for the unknown functions and the spatial lattice information, we reformulate \eqref{Eq:SNR} as
\begin{equation}
\boldsymbol{\eta^{\circ}}=\mathbf{\iota}\boldsymbol{\beta}_0+\sum^p_{j=1}\mathbf{X}_j\boldsymbol{\beta}_j+\mathbf{Z}\gamma+\mathbf{W}\xi
\label{eq:matrix:form}
\end{equation} 
with $\mathbf{\iota}$ as a vector of ones and $\boldsymbol{\beta}_0$ as vector of possible offset parameters. $\mathbf{X}_j\boldsymbol{\beta}_j$ restates the vector of function evaluations $f_j=(f_j(\nu_{1j}),\ldots,f_j(\nu_{nj}))\T$ in form of a matrix product. Here, $\mathbf{X}_j$ is a design matrix and $\boldsymbol{\beta}_j$ is a vector of unknown coefficients. The general form of the prior for the unknown functions and the spatial lattice information can then be expressed as 
\begin{equation}\label{eq:genpriori}
p(\boldsymbol{\beta}_j|\sigma^2_j)\propto\exp\left(-\frac{1}{2\sigma_j^2}\boldsymbol{\beta}^{\mathsf T}_j\mathbf{K}_j\boldsymbol{\beta}_j\right).
\end{equation}
Here, $\sigma^2_j$ is a  variance parameter which controls the trade-off between flexibility and smoothness and  $\mathbf{K}_j$ is a penalty matrix.

As smoothness prior for the metrical covariates we consider P-splines  which has been introduced by \cite{Eilers1996} in a frequentist setting. Here, the underlying assumption is that the unknown function $f_j$ can be approximated by means of a spline of degree $l$ defined on a set of equally space knots $\nu_j^{min} = \zeta_0 < \zeta_1 < \ldots < \zeta_{\tau-1} < \zeta_\tau  = \nu_j^{max}$ within the domain of the covariate $\nu_j$. Such a spline can then be represented as a linear combination of $m=l+r$ $B$-spline basis functions $B_m$
\[
f_j(\nu_j)=\sum^m_{j=1}\boldsymbol{\beta}_{jm}B_m(x_j)
\]
where $\boldsymbol{\beta}_j$ is a vector of unknown regression coefficients.  For a detailed discussion of alternative prior choice we refer the interested reader to \cite{Brezger2006}, \cite{Fahrmeir2004} and \cite{kneib2006}.

For the empirical Bayesian estimation we reparametrize the SNR model in terms of a generalized linear mixed models (GLMM).  
As described in \cite{green1987} for splines and in \cite{Fahrmeir2004} and \cite{kneib2006} for the STAR model, we decompose $\boldsymbol{\beta}_j$ into a penalized ($p$) and a unpenalized part ($q$) such that $\boldsymbol{\beta}_j=\mathbf{X}^{(p)}_j\boldsymbol{\beta}_j^{(p)}+\mathbf{X}^{(q)}_j\boldsymbol{\beta}_j^{(q)}$. From this, we can rewrite $\boldsymbol{\beta}^{\mathsf T}_j\mathbf{K}_j\boldsymbol{\beta}_j$ of \eqref{eq:genpriori} as  $\boldsymbol{\beta}_j^{(p){\mathsf T}}\boldsymbol{\beta}_j^{(p)}$.  Then, setting $\mathbf{X}^{+}_j=\mathbf{X}_j\mathbf{X}^{(p)}_j$ and $\mathbf{X}^{-}_j=\mathbf{X}_j\mathbf{X}^{(q)}_j$ we can rewrite our structured network predictor  as
\[
\eta^{\circ}_j=\mathbf{X}^{+}_j\boldsymbol{\beta}^{(p)}_j+\mathbf{X}^{-}_j\boldsymbol{\beta}^{(q)}_j.
\]

Here, as prior we assume $p(\beta_j^{(q)})\propto c$ and 
 $p(\beta_j^{(p)})\sim N(0,\sigma_j^2I))$.
Using this result, the estimation of the regression and variance parameters can be performed using GLMM techniques, namely iteratively weighted least squares (IWLS) and restricted maximum likelihood (REML).

\section{Application: urban disturbances-related data}\label{sec:apl}
For illustration of the SNR model, we considered the dependence of the nodewise mean intensity of neighbor and community disturbances on a set of various covariates and graph statistics. 

The data was obtained from the local officials of the city of Castell\'{o}n (Spain) and contains the georeferenced coordinates of phone calls received by the police station as well as a set of $32$ additional covariate information. The listed calls were received at the local police call centre or transferred by 112 emergency service to the local police call centre. Geo-codification was  performed indirectly by local officials based on precise address information provided by the caller. The calls comprise up to nine different types of crimes or anti-social behavior. From this data, we pre-selected a sample of  $N=9790$ events classified as neighbor and community disturbances.

The city of Castell\'{o}n is divided into $108$ census sub-areas with an overall surface of $108659 km^2$. According to the information given by the city hall, the total amount of inhabitants is $181616$ of people at the end of 2010. Here, the analysis is based on a subset of phone calls received from the city center that has an overall surface of $8616 Km^2$ divided in $89$ census sub-areas and 130294 inhabitants.

\begin{figure}[H]
\begin{center}
\includegraphics[width=3in]{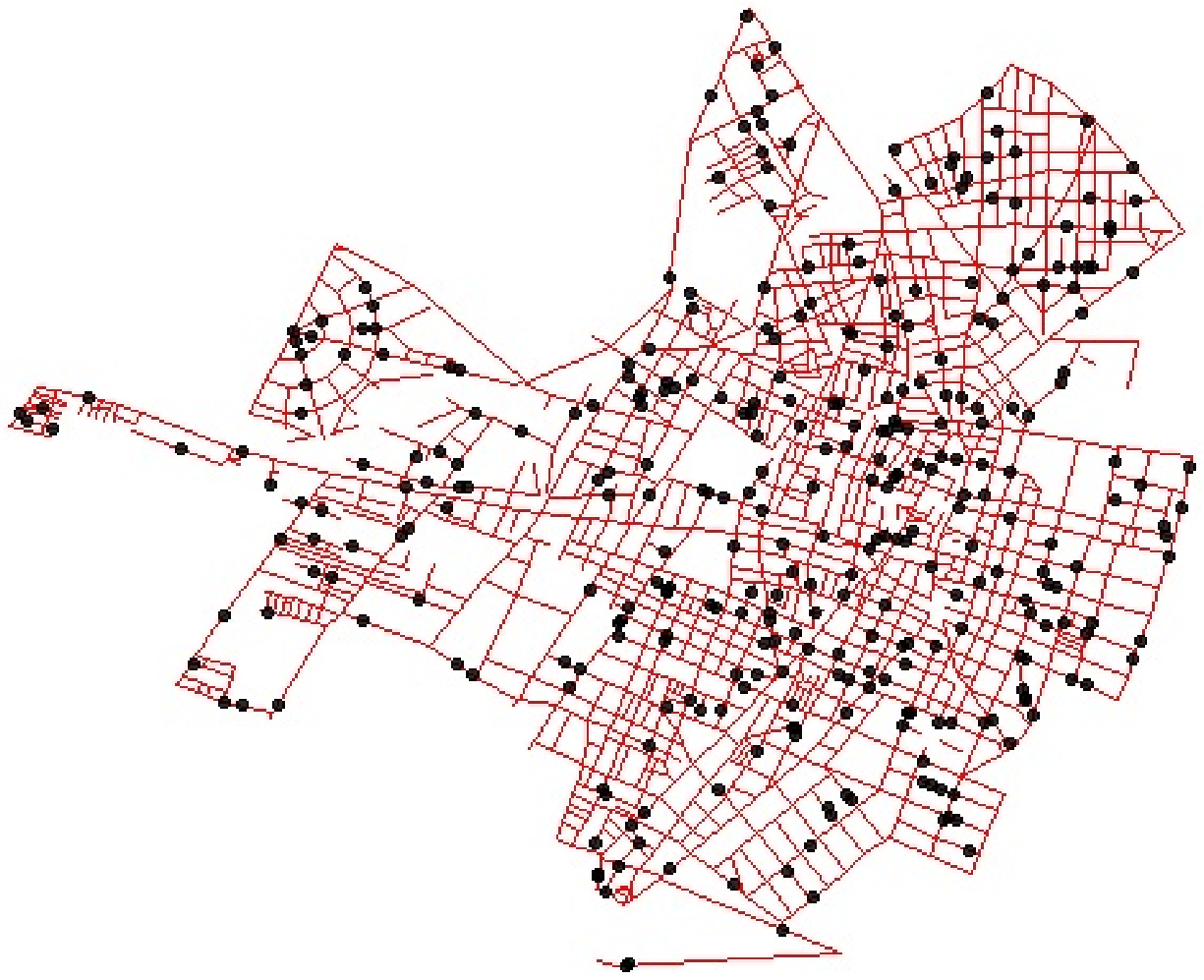}
\caption{Castell\'{o}n traffic network where solid pink lines indicate streets and events are plotted as black dots.   \label{fig:1}}
\end{center}
\end{figure}

For the analysis, we selected in total $1611$ segmenting locations of the traffic network treated as the vertex set of our network graph. So, the vertex set contains $1611$ single nodes, two isolated nodes were excluded. The corresponding traffic network is shown in Figure \ref{fig:1} where we depicted some events as black dots. 

In this graph, $34$ nodes have a degree of one, $30$ nodes a degree of two and $368$ vertices a degree of three. Additionally, we observed four adjacent nodes that have been reported for $181$ locations while a vertex degree of five only appears once. The mean degree in this network is $3.14$. The graph consists of $21$ components. In addition, the length of the longest path is $64$.

To each vertex we attribute the precise georeferenced coordinates of the segmenting location. For any edge in the edge set we calculated the interval length as the squared geodesic distance between pairs of these coordinated vertices. Form this procedure we obtained nodewise mean intensity values for $614$ out of $1611$ vertices. 
Similarly, we calculated nodewise mean values and proportion for all covariates.  Besides, the degree and the betweenness centrality measures of the graph were also used as covariates. The betweenness centrality measure expresses the number of shortest paths passing through a certain node. In our graph we observed a mean betweenness of $18210$. The maximum betweenness was $204000$.

The betweenness centrality measures was also used to detect communities structures in the graph where we recursively extracted the edge with the highest betweenness value \citep[cf.][]{Newman2004}. This results in $45$ disjoint community groups. 

\begin{figure}[H]
\begin{center}
\includegraphics[width=3in]{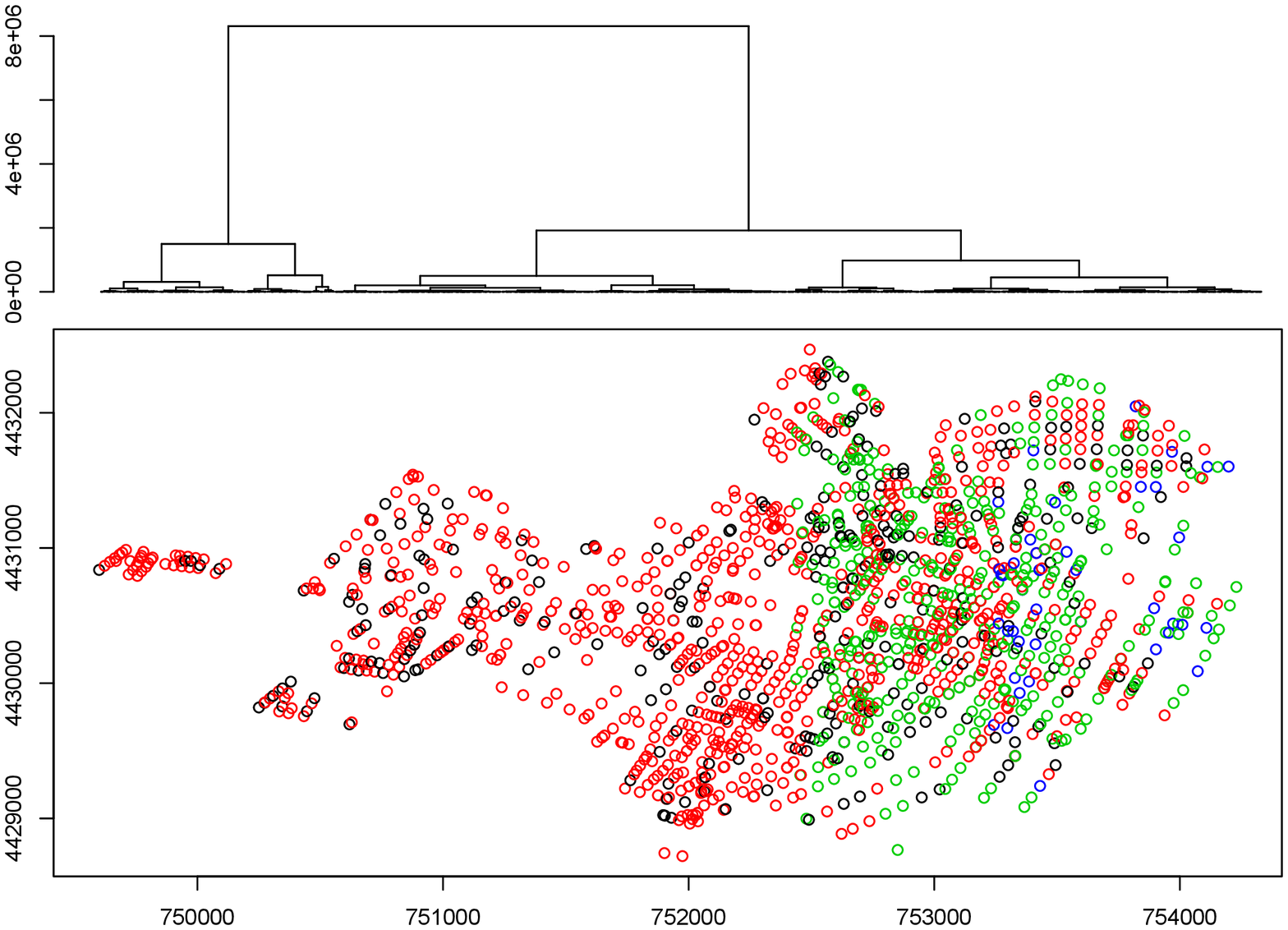}
\end{center}
\caption{Four cluster solution of covariate effects for the complete Castell\'{o}n network. Colour red indicates nodes for which the mean nodewise intensity has been zero. \label{fig:2}}
\end{figure}

In a first step, we perform a hierarchical cluster analysis to detect similarity structures within our data. A four cluster solution using the Ward algorithm is depicted in Figure \ref{fig:2} where the color red highlights nodes which were treated as missing values in our analysis.

Here, we observed that most of the calls which have been classified as neighbor and community disturbances were located in the city centre of Castell\'{o}n.

In a second step, we performed variable selection procedures using a generalized cross-validated Lasso and also classification and regression trees. Based on this analysis, we selected a covariate set of $5$ metric covariates and also the degree statistic. As continuous covariates we included the soil value indicator as well as the closest distances to pharmacies, parks, education and also health centres. The summary statistics of the nodewise means of all continuous covariates measures are reported in Table \ref{tab:1}. 

\begin{table}
\caption{Summary statistics for the covariates of the SNR model. \label{tab:1}}
\begin{center}
\begin{tabular}{rrrrrrr}
          &  Min &   1st Q &  Median     &  Mean       &  3rd Q &  Max \\ \hline
          distance to the closest pharmacy   & $7.62$  & $71.36$  & $107.15$  & $145.17$  & $169.48$    & $859.17$  \\
distance to the closest health center & $36.46$ & $214.41$   & $316.47$   & $350.68$   & $439.77$     & $1161.20$  \\
distance to the closest park   & $0.00$ & $73.52$    & $127.37$   & $146.77$   & $199.86$     & $521.76$  \\
distance to the closest education center  & $0.50$ & $106.80$  & $159.26$  & $185.61$  & $219.77$    & $719.72$  \\
soil value indicator &$171.0$   &$450.0$  & $537.0$  & $551.7$ &  $632.0$  & $946.0$ \\ 
\end{tabular}
\end{center}
\end{table}

\subsection{Results of the SNR model}

We now discuss the results of our final SNR model. For selection, we implemented different SNR models and investigated the Akaike information criterion (AIC), Bayesian information criterion (BIC) and generalized cross-validation statistic (GCV) for model selection between competing models. As covariate set we considered the degree, the distance to the closest pharmacy, the distance to the closest park, the distance to the closest health center, the distance to the closest education center and the soil value indicator. 

To evaluate the performance of the SNR model, we considered four alternative models ($mod1$ - $mod4$). For $mod1$ and $mod2$ we excluded the degree measures from the covariate set. As alternative, $mod3$ and $mod4$ were computed for all $6$ covariates. Additionally, we only considered P-splines  for the distance to the closest park and the soil value indicator in $mod2$ and $mod4$. The AIC, BIC and GCV values for the null model and all four competitive models are reported in Table \ref{tab:one}.    

\begin{table}
\caption{Information criteria and generalized cross-validation for different
models \label{tab:one}}
\begin{center}
\begin{tabular}{rrrr}
 Model         &  AIC &   BIC &  GCV \\ \hline
$null$ & $ 1042.85$  & $1047.27$ & $1.66705$\\
$mod1$ & $876.094$  & $902.614$  & $1.40099$ \\ 
$mod2$ &  $807.698$ &$896.986$  &$1.30024$\\
$mod3$ & $683.808$  &$728.008$  & $1.08252$\\
$mod4$ & $672.84$  &$737.003$   &$1.06472$\\ 
\end{tabular}
\end{center}
\end{table}

Here, we observed that the inclusion of the degree measure improves the model performance. When considering the AIC and the GCV statistics, we found that the inclusion of nonlinear terms in the SNR model leads to an improved fit.

Based on these results, we selected $mod4$ as our final model. Thus, our final model includes the vertex degree and the distance to the closest pharmacy, the distance to  the closest health center and the distance to closest education center were chosen as fixed effects. The results for the fixed effects are reported in Table \ref{tab:2} where we chose a vertex degree of 1 as the reference category. 

\begin{table}
\caption{Parametric coefficients of the SNR model. \label{tab:2}}
\begin{center}
\begin{tabular}{rrrrr}
                 &  Estimate &  Std. Error &  t value & Pr($>|t|$) \\ 
                 \hline   
(Intercept)      & $1.88$   &   $ 0.16$  &$ 12.08$  &   $<2e-16$ \\
node degree = $2$  &$-0.82$   &   $0.15$   &$-5.46 $  &   $<2e-16$   \\
node degree = $3$  &$-1.24$   &   $0.09$   &$-13.9$  &   $<2e-16$  \\
node degree = $4$  &$-1.45$   &   $0.11$   &$-13.39$  &   $<2e-16$  \\
node degree = $5$  &$-3.22$   &   $2.36$   &$ -1.36$  &   $0.1740$  \\ 
distance to the closest pharmacy          &$-0.0019$   &   $0.0004$   &$-4.2171 $  &   $<2e-16$   \\
distance to the closest health center       &$-0.0006$   &   $0.0002$   &$-2.8411 $  &   $0.0046$   \\
 distance to the closest education center         &$ 0.0008$   &   $0.0004$   &$ 2.0211 $  &   $0.0437$   \\ 
\end{tabular}
\end{center}
\end{table}

Except for degree = $5$, we observe significant effects for all covariates.  
For the impact of the degree measure, we observe that the nodewise mean intensity of neighbor and community disturbances decreases if the node degree increases. This might indicate that the subjective perception of neighbor and community disturbances strictly depends on the number of adjacent street segments in the Castell\'{o}n traffic net. This could mean that the subjective threshold of e.g. noise pollution is less strict for inhabitants of highly structured traffic areas, as these areas are commonly expected to be noisier.     

 A similar effect is shown for the nodewise mean distance to the closest pharmacy and the nodewise mean distance to the closest health center. Interestingly, for the nodewise mean distance to the closest education center, a positive effect is reported.

\begin{figure}
\begin{center}
\includegraphics[width=3in]{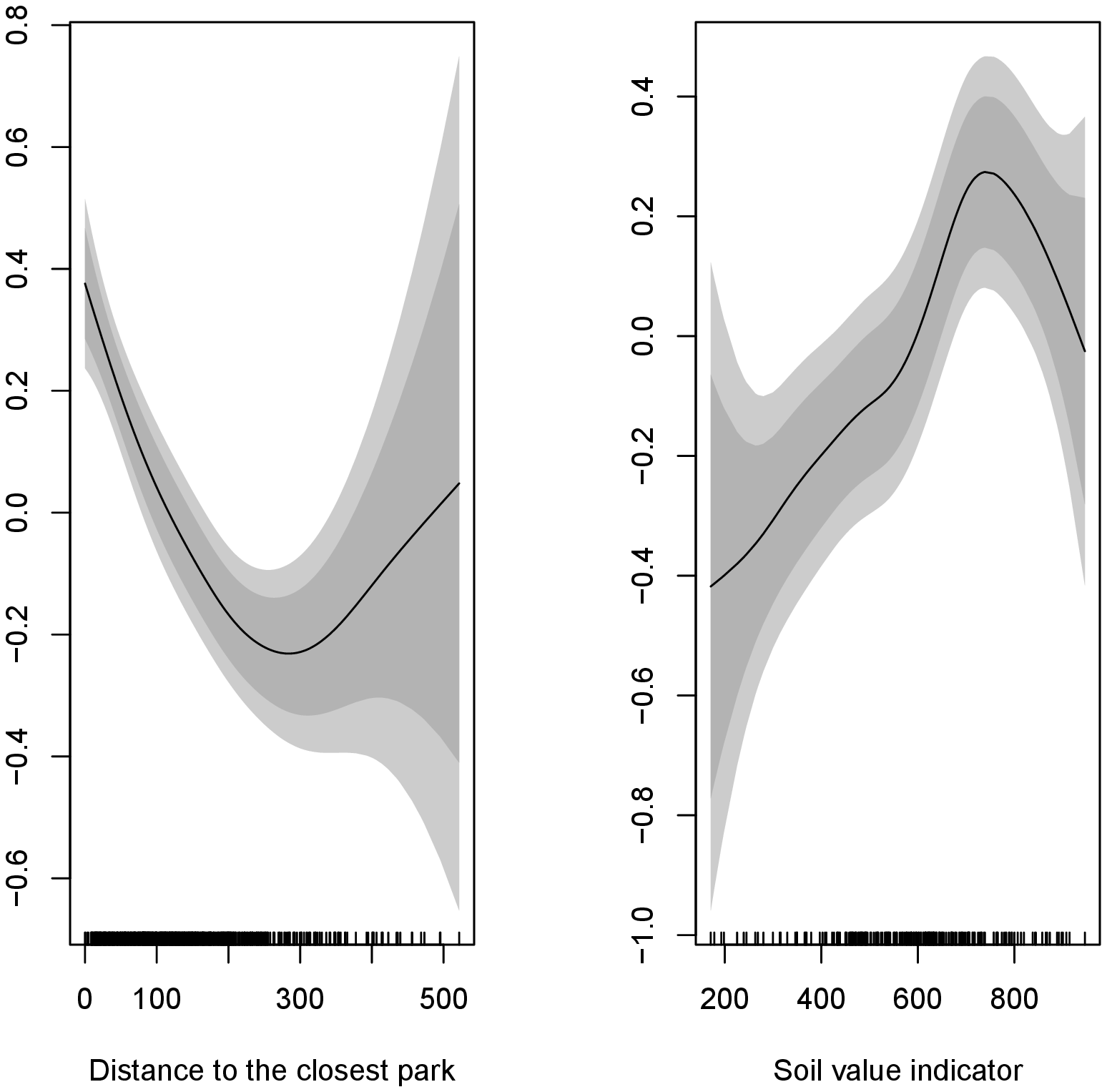}
\end{center}
\caption{Nonparametric effects of the distance to the closest park and of the soil value indicator with pointwise  80\% and 95\% credible intervals. \label{fig:3}}
\end{figure}

In addition, we modeled the effect of the distance to the closest park and the soil value indicator on the nodewise mean intensity of neighbor and community disturbances nonlinearly using P-splines. The estimated nonlinear effects  for the  covariates distance to the closest park and soil value indicator together with 80\% and 95\% credible intervals are visualized in Figure \ref{fig:3}.

Here, for the left panel we observed that the nonlinear effects of the distance to the closest park on our outcome variable followed an U-shape relationship. We found that the nodewise mean intensity of neighbor and community disturbances strongly decreases for distances up to 300 meters. Thereafter, a steadily increase is depicted. 

A different effect is shown in the right panel of Figure \ref{fig:3}. Here, the smooth curve expressed an inverse U-shape impact on the nodewise mean intensity of neighbour and community disturbances. 

\section{Conclusions}\label{sec:conclusion}

In this paper, we proposed a structured network regression model which provides a flexible toolbox for analyzing the impact of a set of different covariates and various graph statistics on generic intensity functions in the context of spatial network structures. This SNR model combines generalized structured additive regression models, graph theoretical statistics and the formalism of point patterns which occur on spatial network structures where generic intensity functions are treated as regressand. By doing so, the SNR model explicitly controls for the network structures which will not be addressed by classical regression techniques for spatial point patterns. Neither the network structure nor the structural relations contained in the graph will be captured by classical point pattern methodology and regression techniques.

The unified framework of the SNR model offers new insights in spatial point patterns that occur on complex domains. Various graph statistics and well-known concepts from network analysis can be chosen and integrated in a general regression framework. On the one hand, this leads to a strong increase of possible impact factors and various regression models.  On the other hand, the unified framework provides several fit characteristics such as AIC, BIC or GCV for the evaluation of competitive models.
  
In general, the structured network predictor can be enlarged to allow for heterogeneity including random effect terms. One possibility would be to relate this random effects to the network structure such as edgewise random effects. Extensions to quantile, expectile or mode SNR models are straightforward. In addition, the estimation of the SNR model is performed using well-established fitting algorithms. Finally, the whole estimation could also been carried out applying a fully Bayesian approach and MCMC-techniques instead of the proposed REML and IWLS procedures.

\bibliographystyle{Chicago}
\bibliography{STARnet}

\begin{thebibliography}{}

\bibitem[\protect\citeauthoryear{Ang}{Ang}{2010}]{Ang:2010}
Ang, W. (2010).
\newblock {\em Statistical Methodologies for Events in a Linear Network}.
\newblock Ph.\ D. thesis, The University of Western Australia.

\bibitem[\protect\citeauthoryear{Ang, Baddeley, and Nair}{Ang
  et~al.}{2012}]{Ang:Baddeley:Nair2012}
Ang, W., A.~Baddeley, and G.~Nair (2012).
\newblock Geometrically corrected second order analysis of events on a linear
  network, with applications to ecology and criminology.
\newblock {\em Scandinavian Journal of Statistics\/}~{\em 39}, 591--617.

\bibitem[\protect\citeauthoryear{Baddeley, Jammalamadaka, and Nair}{Baddeley
  et~al.}{2014}]{Baddeley:Jammalamadaka:Nair:2014}
Baddeley, A., Jammalamadaka, and G.~Nair (2014).
\newblock Multitype point process analysis of spines on the dendrite network of
  a neuron.
\newblock {\em Journal of the Royal Statistical Society, Series C (Applied
  Statistics)\/}~{\em 63\/}(5), 673--694.

\bibitem[\protect\citeauthoryear{Borruso}{Borruso}{2005}]{Borruso:2005}
Borruso, G. (2005).
\newblock Network density estimation: Analysis of point patterns over a
  network.
\newblock In O.~Gervasi, M.~Gavrilova, V.~Kumar, A.~Lagan{\'a}, H.~Lee, Y.~Mun,
  D.~Taniar, and C.~Tan (Eds.), {\em Computational Science and its Applications
  - ICCSA}, Number 3482 in Lecture Notes in Computer Science, pp.\  126--132.
  Springer.

\bibitem[\protect\citeauthoryear{Borruso}{Borruso}{2008}]{Borruso:2008}
Borruso, G. (2008).
\newblock Network density estimation: a gis approach for analysing point
  patterns in a network space.
\newblock {\em Transactions in GIS\/}~{\em 12}, 377--402.

\bibitem[\protect\citeauthoryear{Brezger and Lang}{Brezger and
  Lang}{2006}]{Brezger2006}
Brezger, A. and S.~Lang (2006).
\newblock Generalized structured additive regression based on bayesian
  p-splines.
\newblock {\em Computational Statistics \& Data Analysis\/}~{\em 50\/}(4), 967
  -- 991.

\bibitem[\protect\citeauthoryear{Carrington, Scott, and Wasserman}{Carrington
  et~al.}{2005}]{Carrington2005}
Carrington, P.~J., J.~Scott, and S.~Wasserman (2005).
\newblock {\em Models and methods in social network analysis}.
\newblock Structural analysis in the social sciences. Cambridge (U.K.), New
  York (N.Y.): Cambridge University Press.

\bibitem[\protect\citeauthoryear{Eckardt}{Eckardt}{2016}]{Eckardt:2016}
Eckardt, M. (2016).
\newblock Graphical modelling of multivariate spatial point processes.
\newblock Technical report, Department of Computer Science,
  Humboldt-Universit{\"a}t zu Berlin.

\bibitem[\protect\citeauthoryear{Eckardt and Mateu}{Eckardt and
  Mateu}{2016}]{Eckardt:Mateu2016}
Eckardt, M. and J.~Mateu (2016).
\newblock Point patterns occurring on complex structures in space and
  space-time: An alternative network approach.
\newblock Technical report.

\bibitem[\protect\citeauthoryear{Eilers and Marx}{Eilers and
  Marx}{1996}]{Eilers1996}
Eilers, P. H.~C. and B.~D. Marx (1996).
\newblock Flexible smoothing with b-splines and penalties.
\newblock {\em Statistical Science\/}~{\em 11\/}(2), 89--121.

\bibitem[\protect\citeauthoryear{Fahrmeir, Kneib, and Lang}{Fahrmeir
  et~al.}{2004}]{Fahrmeir2004}
Fahrmeir, L., T.~Kneib, and S.~Lang (2004).
\newblock Penalized structured additive regression for spacetime data: a
  bayesian perspective.
\newblock {\em Statistica Sinica\/}, 731--761.

\bibitem[\protect\citeauthoryear{Fahrmeir and Osuna}{Fahrmeir and
  Osuna}{2003}]{Fahrmeir2003}
Fahrmeir, L. and L.~Osuna (2003).
\newblock Structured count data regression.
\newblock Technical report, Department of Statistics,
  Ludwig-Maximilians-Universit\"{a}t M\"{u}nchen.

\bibitem[\protect\citeauthoryear{Fahrmeir and Osuna}{Fahrmeir and
  Osuna}{2006}]{Fahrmeir2006}
Fahrmeir, L. and L.~Osuna (2006).
\newblock Structured additive regression for overdispersed and zero-inflated
  count data.
\newblock {\em Applied Stochastic Models in Business and Industry\/}~{\em
  22\/}(4), 351--369.

\bibitem[\protect\citeauthoryear{Goldenberg, Zheng, Fienberg, and
  Airoldi}{Goldenberg et~al.}{2010}]{Goldenberg:EtAl:2010}
Goldenberg, A., A.~X. Zheng, S.~E. Fienberg, and E.~M. Airoldi (2010).
\newblock A survey of statistical network models.
\newblock {\em Foundations and Trends in Machine Learning\/}~{\em 2\/}(2),
  129--233.

\bibitem[\protect\citeauthoryear{Green}{Green}{1987}]{green1987}
Green, P.~J. (1987).
\newblock Penalized likelihood for general semi-parametric regression models.
\newblock {\em International Statistical Review\/}~{\em 55\/}(3), 245--259.

\bibitem[\protect\citeauthoryear{Hastie and Tibshirani}{Hastie and
  Tibshirani}{1990}]{Hastie1990}
Hastie, T.~J. and R.~J. Tibshirani (1990).
\newblock {\em Generalized additive models}.
\newblock Monographs on statistics and applied probability. London: Chapman \&
  Hall.

\bibitem[\protect\citeauthoryear{Hennerfeind, Brezger, and
  Fahrmeir}{Hennerfeind et~al.}{2006}]{Hennerfeld2006}
Hennerfeind, A., A.~Brezger, and L.~Fahrmeir (2006).
\newblock Geoadditive survival models.
\newblock {\em Journal of the American Statistical Association\/}~{\em
  101\/}(475), 1065--1075.

\bibitem[\protect\citeauthoryear{Kammann and Wand}{Kammann and
  Wand}{2003}]{Kammann2003}
Kammann, E.~E. and M.~P. Wand (2003).
\newblock Geoadditive models.
\newblock {\em Journal of the Royal Statistical Society: Series C (Applied
  Statistics)\/}~{\em 52\/}(1), 1--18.

\bibitem[\protect\citeauthoryear{Kneib}{Kneib}{2006}]{kneib2006}
Kneib, T. (2006).
\newblock {\em Mixed model based inference in structured additive regression}.
\newblock Ph.\ D. thesis, Ludwig-Maximilians-Universit{\"a}t M{\"u}nchen.

\bibitem[\protect\citeauthoryear{Kolaczyk}{Kolaczyk}{2009}]{Kolaczyk:2009}
Kolaczyk, E. (2009).
\newblock {\em Statistical Analysis of Network Data: Methods and Models.}
\newblock Springer.

\bibitem[\protect\citeauthoryear{Koskinen, Lusher, and Robins}{Koskinen
  et~al.}{2011}]{Koskinen:Lusher:Robins:2011}
Koskinen, J., D.~Lusher, and G.~Robins (2011).
\newblock {\em Exponential Random Graph Models.}
\newblock Cambridge University Press, Cambridge.

\bibitem[\protect\citeauthoryear{McCullagh and Nelder}{McCullagh and
  Nelder}{1989}]{McCullagh1989}
McCullagh, P. and J.~A. Nelder (1989).
\newblock {\em Generalized Linear Models}.
\newblock London: Chapman \& Hall / CRC.

\bibitem[\protect\citeauthoryear{Newman and Girvan}{Newman and
  Girvan}{2004}]{Newman2004}
Newman, M. E.~J. and M.~Girvan (2004).
\newblock Finding and evaluating community structure in networks.
\newblock {\em Physical Review E\/}~{\em 69}, 026113.

\bibitem[\protect\citeauthoryear{Okabe and Satoh}{Okabe and
  Satoh}{2009}]{Okabe:Satoh:2009}
Okabe, A. and T.~Satoh (2009).
\newblock Spatial analysis on a network.
\newblock In A.~Fotheringham and P.~Rogers (Eds.), {\em The SAGE Handbook on
  Spatial Analysis}, Chapter~23, pp.\  443--464. SAGE Publications.

\bibitem[\protect\citeauthoryear{Okabe and Sugihara}{Okabe and
  Sugihara}{2012}]{Okabe:Sugihara:2012}
Okabe, A. and K.~Sugihara (2012).
\newblock {\em Spatial Analysis Along Networks}.
\newblock Wiley.

\bibitem[\protect\citeauthoryear{Okabe and Yamada}{Okabe and
  Yamada}{2001}]{Okabe:Yamada:2001}
Okabe, A. and I.~Yamada (2001).
\newblock The ${K}$-function on a network and its computational implementation.
\newblock {\em Geographical Analysis\/}~{\em 33\/}(3), 271--290.

\bibitem[\protect\citeauthoryear{Ripley}{Ripley}{1976}]{ripley:76}
Ripley, B.~D. (1976).
\newblock The second-order analysis of stationary point processes.
\newblock {\em Journal of Applied Probability\/}~{\em 13}, 255--266.

\end{thebibliography}

\end{document}